# Using Ethernet or A Wireless Harness and Named Data Networking in Autonomous Tractor-Trailer Communication

Ahmed Elhadeedy and Jeremy Daily
Colorado State University

## Abstract

Autonomous truck and trailer configurations face challenges when operating in reverse due to the lack of sensing on the trailer. It is anticipated that sensor packages will be installed on existing trailers to extend autonomous operations while operating in reverse in uncontrolled environments, like a customer's loading dock. Power Line Communication (PLC) between the trailer and the tractor cannot support high bandwidth and low latency communication. This paper explores the impact of using Ethernet or a wireless medium for commercial trailer-tractor communication on the lifecycle and operation of trailer electronic control units (ECUs) from a Systems Engineering perspective to address system requirements, integration, and security. Additionally, content-based and host-based networking approaches for in-vehicle communication, such as Named Data Networking (NDN) and IP-based networking are compared. Implementation, testing and evaluation of prototype trailer ECU communication with the tractor ECUs over Ethernet is shown by transmitting different data types simultaneously. The implementation is tested with two networking approaches, Named Data Networking, and Data Distribution Service (DDS) and the test indicated that NDN over TCP is an efficient approach that is capable of meeting automotive communication requirements. Using Ethernet or a wireless harness and NDN for commercial trailer Anti-Lock Braking System (ABS) ECU provides adequate resources for the operation of autonomous trucks and the expansion of its capabilities, and at the same time significantly reduces the complexities compared to when new features are added to legacy communication systems. Using a wireless medium for tractor-trailer communication will bring new cybersecurity challenges and requirements which requires new development and lifecycle considerations.

## Introduction

Level 4 and 5 [1] Autonomous trucks (AT) are designed to travel long distances without a human driver and the AT is subject to be in situations where it needs to drive in reverse, whether it is a maneuver on a public road or needs to park the trailer in delivery or pickup yards without human intervention. Autonomy sensors are currently being placed on the tractor itself and no sensors being placed on the trailer facing backward which makes autonomous driving in reverse a challenge. Some researchers have addressed autonomous truck reverse driving from an algorithm or vehicle control prospective [2] [3] but not from an ECU integration or systems engineering and the impact it would have on the lifecycle.

Truck native communication protocol between the trailer and the tractor such as Power Line Carrier (PLC) has a limited bitrate of 10 *Kb/s* which makes it unsuitable for autonomy applications, so adding autonomy sensors (e.g., LiDAR, camera or ultrasonic sensors) to the back of the trailer would require an additional trailer ECU for signal processing and communication with the network of the AT, which means having two separate ECUs on the trailer, the native and the newly added ECU. Additionally, AT L4/ and L5 autonomy brings its own communication network, such as Ethernet and CAN FD channels on top of the native tractor network channels such as J1939, PLC or ISO11992, so integrating a second trailer ECU with all of the new and native communication channels will be complex.

In this paper, we will discuss vehicle intra-communication related literature and the impact of using Ethernet or a wireless harness and Named Data Networking (NDN) on the different aspects of the trailer ECU from a systems engineering prospective such as the impact on requirements, and security. A comparison between Named Data Networking, and IP-based Data Distribution Service (DDS) is included, in addition, to a test and evaluation of each approach when transferring multiple data types from one device to three other devices.

## *Related Literature*

In this section, we discuss the related art in vehicle networks and ECU communication architecture, such as Wireless Sensor Networks, Wireless Harness and Wireless CAN, Automotive Protocol Conversion or Replacement with Ethernet, Named Data Networking (NDN) and Tractor-Trailer Communication.

### *A. Wireless Sensor Network*

The concept of wireless sensors network is primarily focused on using a wireless medium to transfer the data from the sensors to an ECU. The ECU itself is wired to the vehicle network and follows the traditional automotive architecture.

Parthasarathy et al. conducted an experiment to evaluate the performance of short-range IEEE 802.15.4-based wireless network on a heavy vehicle between the TPMS sensors and a main and an additional gateway ECUs [4]. Lin et al. evaluated the performance of wireless sensor network under Wi-Fi and Bluetooth interference where the sensors are wirelessly communicating with base stations that are hardwired to the ECU [5]. Potdar and Suyog proposed a zone-based wireless sensor network where ECUs are wired to a gateway that communicates wirelessly with different nodes or sensors [6]. Shaer et al. presented the concept of a wireless blind spot detection and embedded microcontroller using XBee DigiMesh [7].

### *B. Wireless Harness and Wireless Controller Area Network*

The wireless harness concept is focused on replacing the wiring of in-vehicle ECU with a wireless medium such as Bluetooth, Wi-Fi, Ultra-Wideband (UWB) and the automotive 60GHz millimeter-wave. The primary focus is on the measurements, characterization of the wireless signal and the impact of different noise factors on the delay between a transmitter and receiver for in-vehicle communication. The existing art does not cover a wireless communication between the trailer and the tractor. Takayama and Kajiwara evaluated the performance of in-vehicle ECU-to-ECU mesh-networking using UWB-IR for various antenna locations [8] and suggested using ceiling reflection for millimeter-wave wireless harness between two ECUs [9]. Similar studies were conducted on ZigBee [10], IEEE 802.11ad [11] and IEEE 802.15.1 [12] for *in-vehicle* communication with positive results in the presence of interference.

Reddy et al. validated the concept of wireless CAN to Bluetooth gateway to enable wireless CAN transmission from an ECU to a CAN bus [13]. Lun Ng et al. presented the Wireless Controller Area Network (WCAN) using the Token Frame Scheme [14] where the ECUs are connected using a token ring topology and communicating using the CAN principles, however, the proposed solution is different form the standard automotive CAN specifications.

*C. Automotive Protocol Conversion or Replacement with IP-based protocol and Ethernet*

The focus of the Ethernet-related literature is on replacing automotive protocols with Ethernet and Internet Protocol (IP) and on converting automotive protocols from and to Ethernet using IP, where a single protocol will be wrapped in an Ethernet frame (e.g., CAN bus messages or FlexRay) into an Ethernet frame but does not include more than one protocol at a time, therefore, it is lacking multi-protocol data management for a single ECU to support heterogeneous automotive protocols in addition to sensors data over ethernet simultaneously for autonomous tractor trailer application.

Zuo et al. evaluated the concept of CAN/CANFD conversion and transmission to Scalable service-Oriented MiddlewarE over IP (SOME/IP) using a gateway that communicates with another ECU via an ethernet link [15]. The concept doesn't enable the ECU to communicate with the main vehicle bus but to another ADAS ECU and does not take other data formats and protocols into account. Nichițelea and Unguritu proposed using a different Electric and Electronic architecture that is completely based on automotive Ethernet using SOME/IP [16]. Data Distribution Service (DDS) was also proposed for Automotive Software Architectures using IP [17]. Postolache et al. presented an implementation and testing of packing multiple CAN frames in an Ethernet frame using a CAN-Ethernet gateway [18]. Lee et al. also presented a design of a FlexRay/Ethernet Gateway to pack FlexRay messages in Ethernet packets [19]. Kim et al. proposed a gateway framework that supports message routing between two protocols, such as routing and converting messages from CAN to FlexRay or Ethernet. In addition, the gateway is capable of routing of Diagnostics over IP (DoIP) messages to Unified Diagnostic Services (UDS) on CAN or FlexRay, the message translation to another protocol is happening inside the gateway itself using pre-defined routing and translation tables using Automotive open system architecture (AUTOSAR) [20]. Ashjaei et al. addressed and presented an overview of Time-Sensitive Networking (TSN) in automotive applications [21] which includes current and the future trends in vehicle networks such as Domain Controller Unit (DCU) replaces legacy gateways with automotive Ethernet as a backbone for the vehicle network architecture [22] [23] [24]. Audio Video Transport Protocol (AVTP) is used to wrap automotive protocols frames or IEC 61883-compliant multimedia in IEEE 1722 Ethernet frames. This method requires the type of data that will be wrapped in the Ethernet frame to be the same in the Ethernet frame such as all CAN messages, all FlexRay or all IEC 61883-4 (i.e., MPEG2-TS Video) data [25] [26] [27] [28].

Some literature suggests replacing automotive protocols and architectures such as CAN or FlexRay with a completely different topology or networking protocol (i.e., all Ethernet) such as [21] [22] [23]. Kraus et al. proposed the replacement of automotive CAN with optical data communication using an optical bus and central processing unit that manages and monitors all the connected devices using Stream Control Transmission Protocol (SCTP) [29]. Nichițelea and Unguritu proposed replacing standard serial protocols (i.e., CAN, FlexRay and LIN) with ethernet [30] using SOME/IP [31].

*F. Named Data Networking (NDN)*

Automotive NDN literature is mainly focused on the connected vehicles and Vehicle-to-Everything (V2X) communication [32] [33] [34] [35] [36] [37] and there is a limited number of papers that addresses using NDN in intra-vehicle communication and the integration with existing automotive protocols. Papadopoulos et al. presented the concept of using NDN for in-vehicle communication with ECU experimentation as a future step suggesting that NDN is a better network approach [38] and in [39], they presented Name-based secure communication architecture for in-vehicle communication suggesting that NDN is an improved IP alternative. Threet et al. demonstrated secure CAN communication using NDN between two Raspberry Pis with average latency for CAN Interest/Data packets of 73 milliseconds [40]. Some researchers have shown that NDN-based networks have better latency performance than IP-based networks [41] using ndnSIM to simulate internet network that connects multiple cities. The existing art was a motive for us to evaluate the performance of NDN when used in the context of autonomous vehicles with multiple ECUs intra-communicating with CAN and two sensors' data as shown in the test and evaluation section later in this paper.

*G. Security and Tractor-Trailer Communication*

The existing art covers the topic of autonomous trucks driving in-reverse from an algorithm and controls perspective [42] [43] but not from a trailer ECU architecture, integration, or systems-thinking point of view. Non-Autonomous Tractor communication art includes different solutions such as combining CANopen and J1939 networks [44], and securing and encrypting the J1939 and diagnostics traffic on the tractor side as Daily et al. presented in [45] [46]. Power Line Carrier (PLC) is being used as a low-speed communication bus between the trailer and the tractor which is not suitable for AT applications due to the bitrate limitation in PLC, where the preamble bitrate is 8772 bits per second and the data body bitrate is 10,000 bits per second [47]. Recent research has shown that PLC communication is vulnerable to hacking and missing authentication on some critical functions as disclosed by National Motor Freight Traffic Association, Inc. (NMFTA) [48] [49] with countermeasures proposed. Additional autonomous and Heavy-duty Vehicles security vulnerabilities are discussed in [50] [51] [52] [53]. Goers and Kühne presented transmitting CAN and sensors data over automotive Ethernet to the truck Advanced Driver-Assistance System (ADAS) ECU [54]. Their long-term proposed solution is for the trailer ABS to have an Ethernet switch, Microcontroller Unit (MCU), ISO 11992 CAN and a multiplexer that communicates with the tractor over a coiled cable. Extending the ISO 11992 standard to include an additional physical layer such as Ethernet was also mentioned. Technology and Maintenance Council (TMC) presented the need for an automated tractor-trailer coupler and the need for a higher data transmission speed between the trailer and the tractor [55]. They also mentioned controlling the trailer lights (e.g., stop light and turn signals) can be controlled over CAN instead of using separate electrical conductors.

## Architecture

On the basis of the above reflections on the state of the field, we can understand that there is a need for a systems-thinking based solution that addresses the following gaps in the existing art:

1) Using the content-centric NDN protocol as a network protocol instead of the traditional host-centric IP for intra-communication, between the trailer ECU and the tractor.
2) Test, evaluation, and empirical data of NDN when used for autonomous vehicles intra-communication over Ethernet or a wireless medium.
3) Using a wireless medium as the only communication link between the trailer and the tractor and the impact on the lifecycle.
4) Upgrade of the trailer ABS architecture to meet the needs of L4/L5 ATs and adding Telematics, lights control and GPS to the trailer ABS ECU instead of having a separate ECUs or hardware module.

### *Ethernet-Based Trailer ABS*

Similar to the solution presented in [54], we propose an enhanced architecture for the trailer ABS ECU to better fit the needs of the conventional trucks and SAE level 4 and 5 Autonomous Tractor and to combine all of the existing trailer features in one ECU instead of having separate hardware modules such as telematics and GPS. SAE Level 4 and 5 autonomy will bring additional communication buses to the tractor to meeting timing and bandwidth requirements such as adding CAN FD and Ethernet on top of tractor platform and the trailer ECU is expected to communicate with the AT over these channels since it will be part of the autonomy hardware feeding sensors data to the self-driving software. Retrofitting an additional new trailer ECU will be complex and result in a big harness as shown in figure 1. The proposed new trailer ABS architecture differs from the existing art as follows: adding telematics and GPS within the ABS, replacing the MCU with a SoC, removing the physical CAN from the trailer ECU, using a multi-Gig Ethernet and adding several interfaces to the MCU at the tractor side as shown in figure 1.

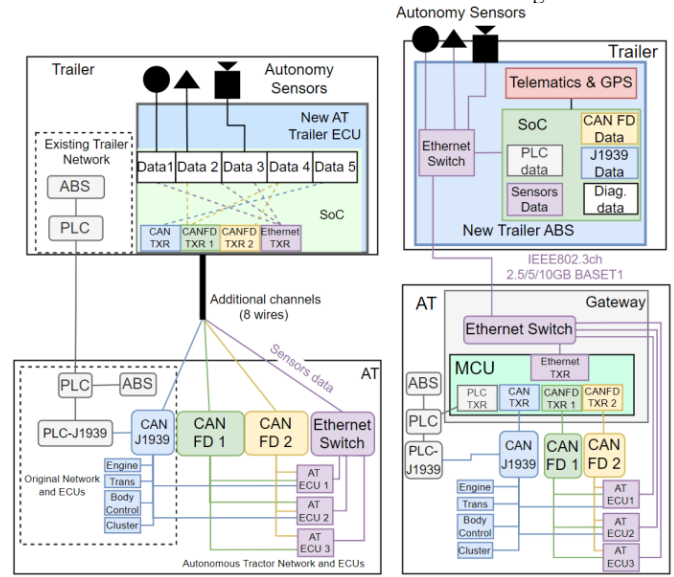

Figure 1. (Left) Retrofitting an autonomy ECU in the trailer to process rear sensors to accommodate level 4 and 5 autonomy needs. (Right) Proposed upgraded trailer ABS architecture that integrates with the existing autonomous tractor architecture.

### *Wireless Trailer ABS*

We propose replacing the physical data cables between the trailer and the tractor with a wireless medium. Wireless Trailer ABS architecture will be similar to the Ethernet-based architecture except for the physical link between the trailer and the tractor. Wireless harness concept for *in-vehicle* ECU communication has been previously proposed and tested using different wireless mediums such as Ultra-wideband [8], millimeter-waves [9], and 60 GHz Wi-Fi [11] with positive results such as achieving a data rate of 600 to 700 *Mbps* in the case of the using IEEE 802.11ad. The wireless harness has not been addressed or proposed when used between the trailer and the tractor as a communication link. The wireless harness (e.g., IEEE 802.11ad) supports multi-Gig data rate and could be leveraged in the communication between the tractor and trailer ABS ECU to enable automatic pairing and eliminate the need for plugging a cable for conventional trucks or autonomous trucks, especially if the process of coupling and uncoupling a trailer and a tractor needs to be automated without human intervention. The wireless trailer ABS can contain any functionalities needed on the trailer side such as braking, traction control, trailer monitoring and diagnostics, lights control, telematics, or any additional functionalities to reduce the number of ECUs or hardware modules needed. The wireless harness using 60 GHz Wi-Fi, for example, could have a standardized range for truck-trailer communication only to avoid interference with other applications.

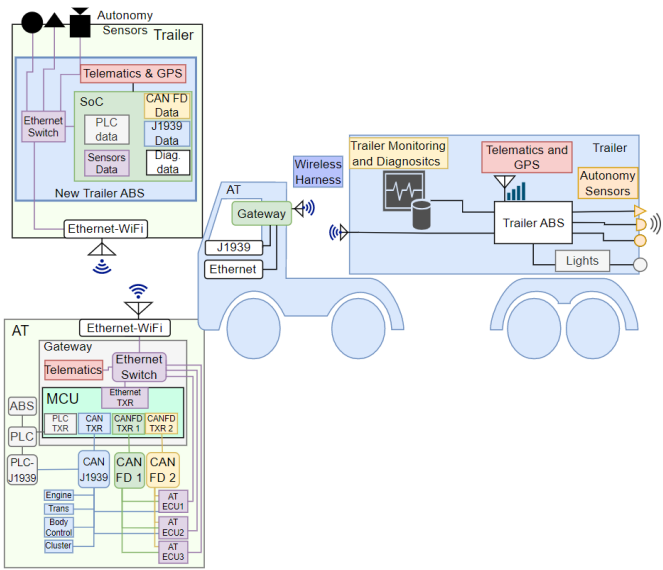

Figure 2.  Proposed wireless trailer ABS ECU using a wireless harness and with more features combined such as telematics, GPS, and sensors data processing.

As discussed in the related literature section, future in-vehicle network includes DCU and Ethernet as a backbone, so for conventional trucks, a future network concept is shown in figure 3.

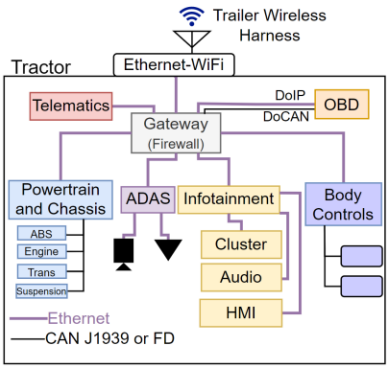

Figure 3.  Future network architecture concept for trucks

## System Requirements

In order for the new trailer ABS to accommodate the autonomy needs, it needs to adhere to the following requirements:

Table 1. New Ethernet-based trailer ABS ECU system requirements

| | |
|---|---|
| R1 | The trailer ABS ECU shall have additional inputs for different types of sensors such as the LiDAR, camera or ultrasonic to support autonomous driving in reverse. |
| R2 | The sensors shall be mounted on the back of the trailer facing backwards to function as the reverse sensing system for the autonomous tractor. |
| R3 | The trailer ECU shall be able to process sensors data and transmit it to the tractor. |
| R4 | The trailer ECU shall only use Ethernet (e.g., IEEE 802.3ch 2.5/5/10GB BASET1) to communicate with the tractor. |
| R5 | The trailer ECU shall be able to wrap multiple communication protocols over Ethernet such as sensors data, AT CAN FD, CAN buses, tractor standard J1939 CAN, or diagnostics data simultaneously. |
| R6 | A gateway shall be added on the AT side as an extension to the trailer ECU to route the incoming traffic to the appropriate buses at the AT. Similarly, the gateway shall wrap the traffic going from the AT to the trailer over Ethernet. |
| R7 | The system shall integrate and be compatible with legacy truck communication protocols |
| R8 | All the trailer features shall be included in the trailer ABS with no separate hardware modules. Features such as Telematics, GPS, temperature monitoring, trailer monitoring, lights control, Tire Pressure Monitoring System (TPMS), sensors data processing and communication over Ethernet. |
| R9 | The system shall support data authenticity and confidentiality |
| R10 | The system shall be able to meet the data timing requirements (e.g., 20 milliseconds for some CAN messages) |
| R11 | The system shall be interoperable and compatible with tractors from different manufacturers |
| R12 | The system shall operate with many tractors within the same fleet and support one-to-many relationship without conflicts or mixing information. |

In case of a wireless medium between the trailer and the tractor:

Table 2. New wireless harness-based trailer ABS ECU system requirements

| | |
|---|---|
| RW1 | A multi-gig wireless medium shall be used for communication between the trailer and the AT |
| RW2 | The trailer ECU shall be able to exchange different data types including sensors and CAN data over the wireless medium with the AT |
| RW3 | The trailer and the AT shall be able to automatically pair upon power up and being physically connected. |
| RW4 | Each of the trailer and the tractor shall be authenticated by a mutual root of trust before establishing a connection and being paired |
| RW5 | After establishing a connection, the trailer and the tractor shall be able to exchange public keys for data packets encryption and authentication |
| RW6 | The trailer wireless connection antenna (e.g., IEEE 802.11ad) shall be mounted on the front side of the trailer facing the back of the tractor where the tractor antenna is mounted. |
| RW7 | All the data traffic exchanged between the ECUs shall be authenticated |

## Networking Protocol

Using Ethernet as the only communication link between the trailer and the autonomous tractor will require a proper networking protocol to allow the ECUs on the same network to exchange data, meet the timing requirements and support data authenticity. In this section, we will compare two different networking approaches, Named Data Networking (NDN) and Data Distribution Service (DDS).

### *Comparison Between NDN and DDS*

Named Data Networking [56] is a content-based architecture that uses names for each data type, the data transfer is driven by the receiving device where it sends an interest packet containing the name of data and sender responds with a data packet that contains the name, the requested data and the signature as shown in Figure 4. NDN changes the architecture from an Internet Protocol (IP) address-based to a name-based, which makes it simpler to configure the network. Named Data Networking Forwarding Daemon (NFD) [57] is the network forwarder responsible for forwarding interest packets and data packets in each device. NFD forwards and communicates the packets over multiple network interfaces (Face), physical such as Ethernet, transport layers such as Transmission Control Protocol (TCP) and User Datagram Protocol (UDP) or inter-process channel between the NFD and an application. A Face is also the interface connecting two devices on the network with an ID, local Uniform Resource Identifier (URI) and a remote URI for the remote device. NDN supports data security [58] by default and the signature is built-in the data packet structure.

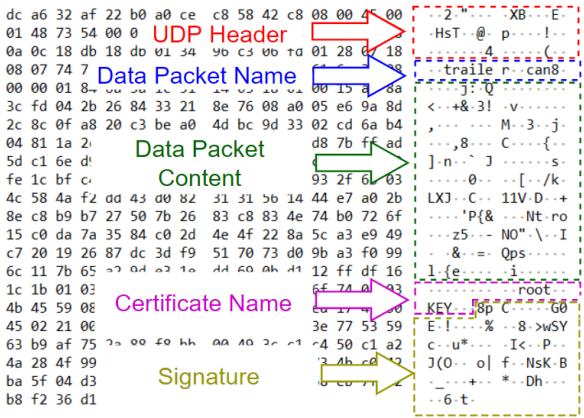

Figure 4. A demonstration of how a NDN data packet named /trailer/can looks over Ethernet.

DDS Real-Time Publish Subscribe (RTPS) is a known protocol that is used in high performance, low latency and real-time communication using the publisher-subscriber architecture.

## Test and Evaluation

The purpose of this test is to evaluate the performance of NDN compared to a well-established and high-performing protocol such as DDS Real-Time Publish Subscribe (RTPS) using the same setup and logic. In this test, NDN over TCP, NDN over UDP and DDS RTPS over UDP are compared, each in a separate run using the same data types, lengths, and timings. For the purpose of testing the baseline performance of each approach, data packet signing at the transmitter and validation at the receiver were not included.

## Test Setup

The test setup included a PC that represents the trailer ECU transferring 3 data types to 3 Raspberry Pis (RPi) connected over an Ethernet switch as shown in Figure 5. The data types are CAN, LiDAR and camera packets. Python scripts were using NDN [59] which is completely implemented using Python and DDS [60] which contains a Python API for an underlying C++ library. The PC is the transmitter that hosted 3 scripts running simultaneously, 1 script for LiDAR data transmitting random 2496 bytes every 1 milliseconds to RPi 1, 1 script for CAN data transmitting random 160 bytes (i.e., 4 CAN and 2 CAN FD signals payload) every 8 milliseconds to RPi 2 and 1 script for camera data transmitting random 8000 bytes every 20 milliseconds to RPi 3. Each of the RPis hosted one receiving script to receive one data type from the PC as shown in Figure 6. The data generated and transmitted to the receiver "D" is strings in some runs and bytes in other runs to compare encoding performance which will be specified in the results.

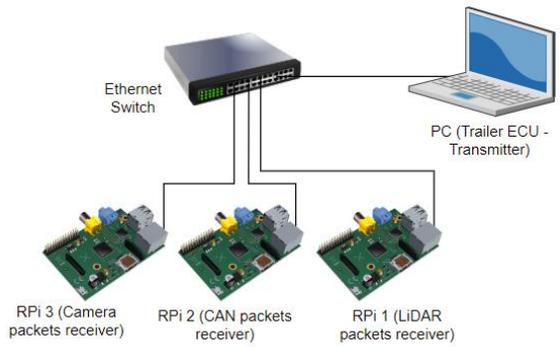

Figure 5. Test setup used to test and evaluate 3 networking approaches.

For NDN, each data type is given a name, /trailer/can for CAN data, /trailer/lidar for LiDAR data and /trailer/cam for camera bytes and since each of the RPis is the receiver of one type of data, they send the interest packets using the name to the PC simultaneously and the PC responds respectively to each device with the data packet under the requested name. Similarly for DDS, each script on the PC was configured as a data writer publishing under a topic that represents a data type and the data reader (i.e., subscriber) for the same topic on the RPi reads the data published in that topic. The topics are Lidar, CAN and Cam. RPi subscribed to the lidar topic, RPi 2 subscribed to the CAN topic and RPi 3 subscribed to the Cam topic. The test cases were in three separate runs, each run using one protocol at a time for 10 minutes.

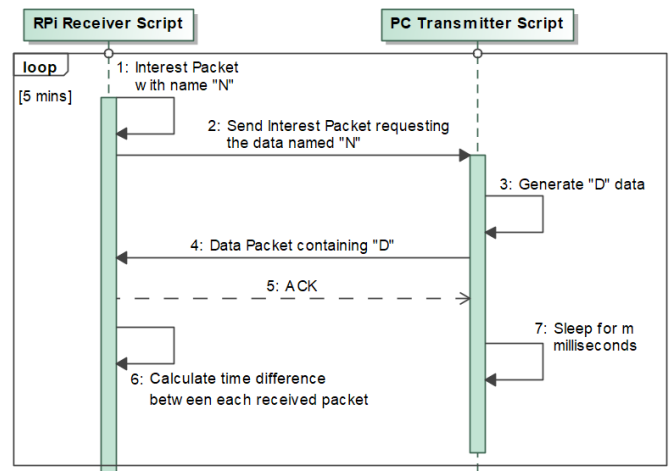

Figure 6. Logic used between the receiver script in each RPi and the PC (transmitter) in the case of NDN over TCP.

## Test Results and Discussion

The results are split into 3 parts: latency performance, core CPU consumption percentage and memory consumption percentage for each of NDN TCP (bytes), NDN UDP (bytes) and NDN TCP, NDN UDP and DDS RTPS using strings. Latency performance was measured in milliseconds (*ms*) and it's the time difference between every received data packet at each receiver (i.e., RPi), in addition to the delay added manually (1 *ms* for Lidar data, 8 *ms* for CAN data and 20 *ms* for cam data). For example, for NDN TCP, each data point represents the time taken for the receiver to send an interest packet, the time taken for the PC to respond with a data packet, time taken for the receiver to respond with an acknowledgement (ACK) and the manually added delay.

*Latency Performance*

In the case of the Lidar data, the PC was set to transmit 2496 bytes or strings representing the lidar data packet every 1 millisecond (hardcoded delay). Table 3 and Figure 7 shows that NDN TCP (bytes) had a similar performance to DDS RTPS where DDS is 50 microseconds ($\mu s$) ahead and NDN had a lower standard deviation of 50 $\mu s$. NDN while sending strings had taken longer due the extra time taken to encode the strings to UTF-8 and packing them into Type-Length-Value formatted packet at both the transmitter and unpacking and decoding at the receiver.

Table 3. Summary of Lidar data transmission latency performance

| Lidar Data ($ms$) | NDN TCP | NDN UDP | NDN TCP | NDN UDP | DDS RTPS |
|---|---|---|---|---|---|
| Data Type | Bytes | | Strings | | |
| Packets Count | 120,023 | 84,477 | 65,807 | 66,465 | 120,932 |
| Mean | 2.50 | 3.75 | 4.6 | 4.5 | 2.45 |
| Latency (mean – 1) | 1.5 | 2.75 | 3.6 | 3.5 | 1.45 |
| Min | 2.08 | 2.09 | 2.8 | 2.8 | 0.16 |
| Max | 6.14 | 7.18 | 11.22 | 11.45 | 8.28 |
| Std. Deviation | 0.45 | 0.52 | 1.13 | 1.07 | 0.50 |

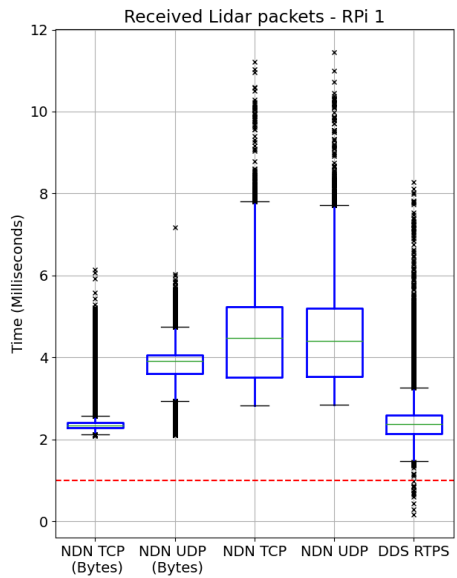

Figure 7. Latency performance of each networking approach when sending Lidar packets.

In the case of the CAN data, the PC was set to transmit 160 bytes or strings representing payload of 4 CAN and 2 CAN FD signals every 8 $ms$ (hardcoded delay). Table 4 and Figure 8 shows similar performance to the previous case with DDS being ahead with 100 $\mu s$ and had a lower standard deviation of 170 $\mu s$.

Table 4. Summary of CAN data transmission latency

| CAN data ($ms$) | NDN TCP | NDN UDP | NDN TCP | NDN UDP | DDS RTPS |
|---|---|---|---|---|---|
| Data Type | Bytes | | Strings | | |
| Packets Count | 35,420 | 32,469 | 33,560 | 33,607 | 35,413 |
| Mean | 8.47 | 9.24 | 8.94 | 8.92 | 8.37 |
| Latency (mean – 8) | 0.47 | 1.24 | 0.94 | 0.92 | 0.37 |
| Min | 6.26 | 7.40 | 7.77 | 7.80 | 6.72 |
| Max | 11.39 | 11.42 | 10.63 | 10.54 | 10.19 |
| Std. Deviation | 0.33 | 0.30 | 0.35 | 0.33 | 0.16 |

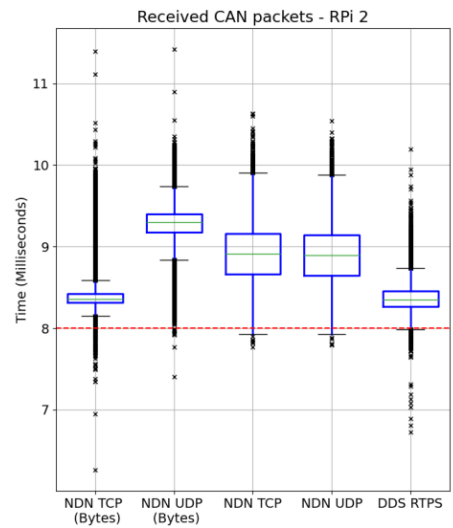

Figure 8. Latency performance of each networking approach when sending CAN packets.

In the case of the camera data, the PC was set to transmit 8000 bytes or strings representing camera data every 20 $ms$ (hardcoded delay). Table 5 and Figure 9 shows that NDN TCP and NDN UDP in the case of bytes had a better performance than DDS with a mean latency of 0.90 $ms$, 2.47 $ms$ and 4 $ms$ respectively which indicates the high efficiency of the NDN over TCP when sending bigger payload of bytes.

Table 5. Summary of camera data transmission latency performance

| Cam data ($ms$) | NDN TCP | NDN UDP | NDN TCP | NDN UDP | DDS RTPS |
|---|---|---|---|---|---|
| Data Type | Bytes | | Strings | | |
| Packets Count | 14,353 | 14,037 | 11,402 | 11,527 | 12,349 |
| Mean | 20.90 | 22.47 | 26.30 | 26.02 | 24.00 |
| Latency (mean – 20) | 0.90 | 2.47 | 6.30 | 6.02 | 4.00 |
| Min | 18.69 | 19.38 | 20.43 | 21.61 | 21.86 |
| Max | 25.68 | 24.61 | 37.54 | 38.53 | 38.58 |
| Std. Deviation | 0.62 | 0.68 | 2.62 | 2.53 | 1.68 |

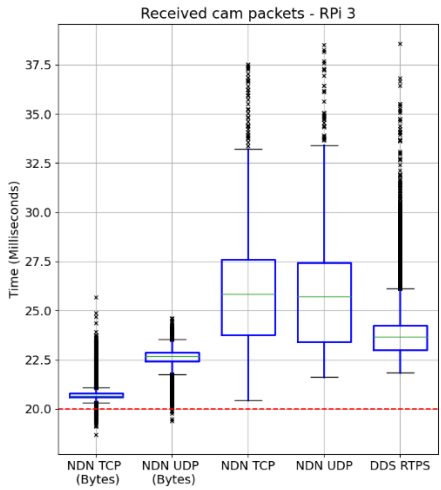

Figure 9. Latency performance of each networking approach when sending camera packets.

*Core CPU Consumption Percentage*

Another performance parameter to investigate when it comes to a new ECU architecture and networking protocol is the CPU consumption to flag resources issues. The core CPU consumption percentage was captured from the Operating System (OS) of each device, the PC, RPi 1, RPi2 and RPi 3. Figure 10 shows the core CPU consumption percentage in the PC while transmitting the three types of data simultaneously in the case of sending strings. Figure 10 includes the power core CPU consumption at the receiver in the three test cases: DDS RTPS, NDN UDP and then NDN TCP. Each test case includes 1 protocol and 3 scripts in the PC transmitting data. Overall, the CPU consumption for both approaches has the same profile with DDS being higher in the case of transmitting Lidar data and NDN being higher in the case of CAN and camera data.

Table 6. Summary of PC (the transmitter) CPU consumption percentage for per script per each protocol

| Mean CPU % | DDS | NDN over UDP | NDN over TCP | NDN over UDP (Bytes) | NDN over TCP (Bytes) |
|---|---|---|---|---|---|
| Lidar script | 50.7% | 44.4% | 44.6% | 33.6% | 32.8% |
| CAN script | 3.7% | 8.8% | 8.6% | 12.3% | 11.6% |
| Cam script | 15.2% | 20.8% | 20.9% | 5.9% | 5.6% |
| NFD | - | 4.4% | 5.8% | 8.3% | 13.1% |

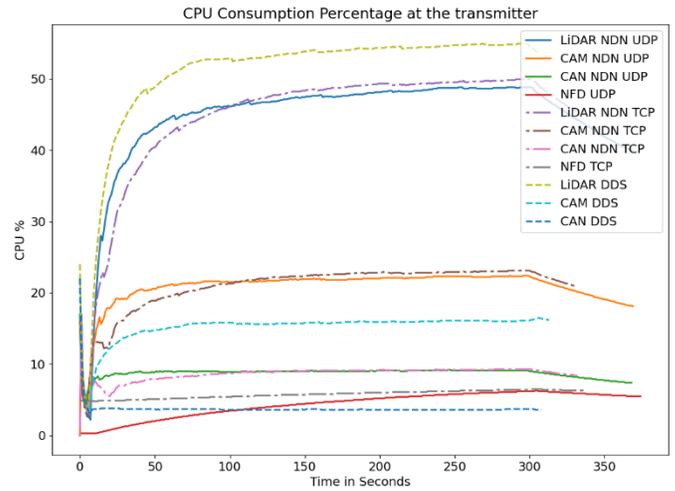

Figure 10. Core CPU Consumption Percentage in the PC while transmitting data using each approach separately when sending strings.

Figure 11 shows the core CPU consumption percentage at each RPi (the receiver) for each protcol. The percentage represrents the consumption from one core CPU only within the Raspberry Pi. The CPU consumption is close in the case of CAN where NDN is higher then DDS by 8% and for camera data DDS was higher than NDN by 10%. For the case of the Lidar, NDN TCP had the highest CPU concumption with a mean of 65.15%, NDN UDP with a mean of 43.96% and DDS had the lowest percentage of 26.23%. In the case of NDN over TCP, the receiver sends an interest packet and an acknowledgement signal every 1 *ms* which explains the high percentage.

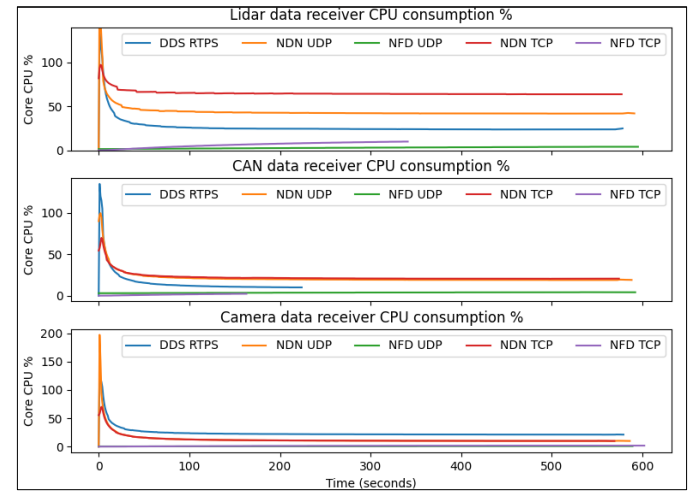

Figure 11. Core CPU Consumption Percentage in the each of the receivers while receiving data using each approach separately.

Memory consumption percentage is shown in Figure 12 and all of the protocols had a similar consumption percentage at about 5% of the available memory.

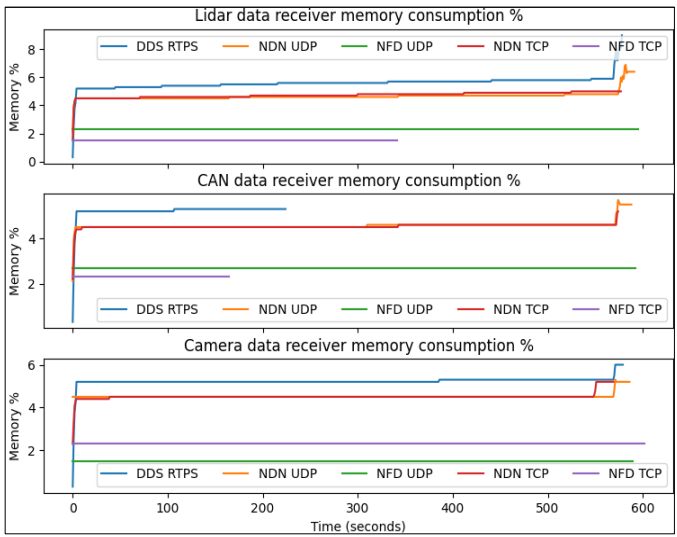

Figure 12. Memory Consumption Percentage in the each of the receivers while receiving data using each approach separately.

In summary, Named Data Networking over TCP performed very well when tested in the same conditions as DDS. NDN over TCP appears to be a very efficient networking approach, especially for bigger payload. NDN is currently under development and the implementation used in this testbed is the general implementation planned for internet architecture, which makes it a strong candidate for automotive networking where it could be tailored and tuned further for automotive applications.

## The Impact of The New System Requirements

Current conventional trailer ECU architecture cannot support autonomy needs for reverse driving due to the design limitation from compute and networking perspectives. Additionally, other trailer features such as telematics, GPS and parking sensing are included in a separate hardware module, therefore, a new architecture is needed for AT as discussed in the requirements in table 1 and table 2. Using Ethernet as the communication link between the trailer and the tractor will have an impact on different aspects of the ECU, which will be addressed in this section.

### The Impact on the Architecture

In R1 from Table 1, adding additional inputs to the ABS ECU to take autonomy sensors data will increase the total number of sensors inputs since the main input from existing trailer sensors is the wheel speed sensors and other trailer monitoring sensors. Processing autonomy sensors requirement as in R3 will result in an increase in the capabilities of the trailer ECU processing such as using a system-on-chip (SoC) instead of what being used currently in conventional trucks such as a Microcontroller (MCU) since the new ABS is expected to do additional data processing before transmitting the data to the AT ECUs.

Using Ethernet or the wireless harness as the only communication link between the trailer and the tractor will result in replacing legacy data communication wires (e.g., PLC, or ISO11992) with automotive Ethernet. The information the legacy communication protocols carry will be digitized and transmitted over Ethernet. To support native protocols and integrate with the tractor side, a MCU is needed to be added on the tractor to function as a gateway for the trailer ECU where it routes the ethernet traffic coming from the new trailer ABS to different buses on the tractor such as PLC, CAN J1939, CAN FD and Ethernet. Additionally, all the features that are being used in the trailer needs to be combined in one ECU to avoid the need for separate hardware modules on the trailer side.

### The Impact on Security

Since the trailer will be a source of information and sensors data for the AT to drive, authentication or encryption of data packets exchanged between the trailer and the tractor will be required. Moreover, in the case of a wireless trailer ABS, an additional authentication step is required to assure that both of the trailer and tractor are authenticated and can connect securely over the wireless harness (e.g., 60 GHz Wi-Fi), especially if the pairing process and coupling the tractor and the trailer is expected to happen without human intervention.

Figure 13 shows Trailer ABS keys provisioning using a cloud-based Fleet Management System (FMS) and a Certificate Authority (CA) over a Virtual Private Network (VPN). FMS will be responsible for managing provisioning, key generation requests, access control and identities of the ECUs and other users. The trailer ABS is shown as an example and the tractor will follow the same process.

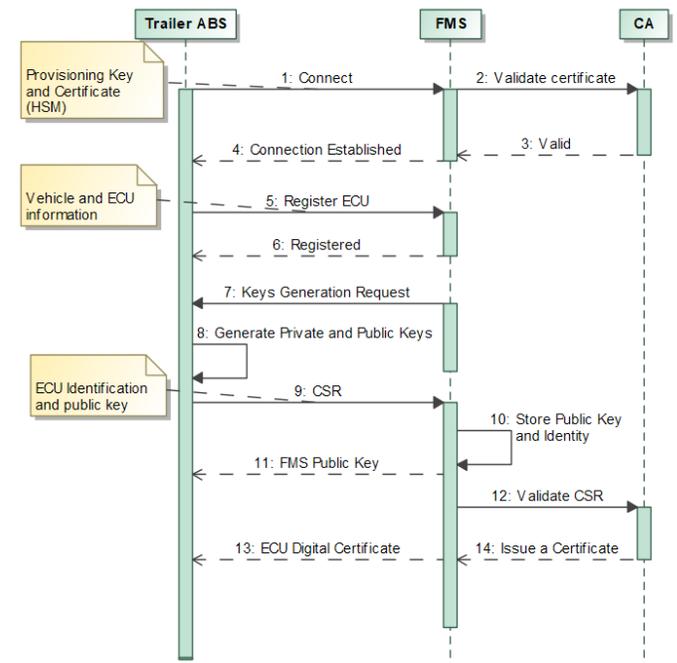

Figure 13. Trailer ECU provisioning using a cloud-based Fleet Management System (FMS).

A method to authenticate the trailer and the tractor using FMS before pairing with each other is shown in figure 14. This is possible by leveraging the telematics unit available in the trailer and the tractor (i.e., cellular connection), in addition to using the GPS coordinates or One-Time Passcode (OTP) as an additional step to verify the integrity of the connection requests. The AT and the trailer will send the connection requests to FMS and FMS will first authenticate each entity. After authentication, GPS coordinates or OPT will be compared and if they matched, the requests can be approved and FMS then will generate Wi-Fi credentials with a defined access level and request the vehicle Wi-Fi server (e.g., AT) to locally update the Wi-Fi credentials and the access list with privileges defined such as the level of access and the types of data supported. After confirmation from the Wi-Fi server that it updated the Wi-Fi credentials and access list, FMS will send the pairing credentials to the Wi-Fi client (e.g., Trailer) to start pairing with the AT.

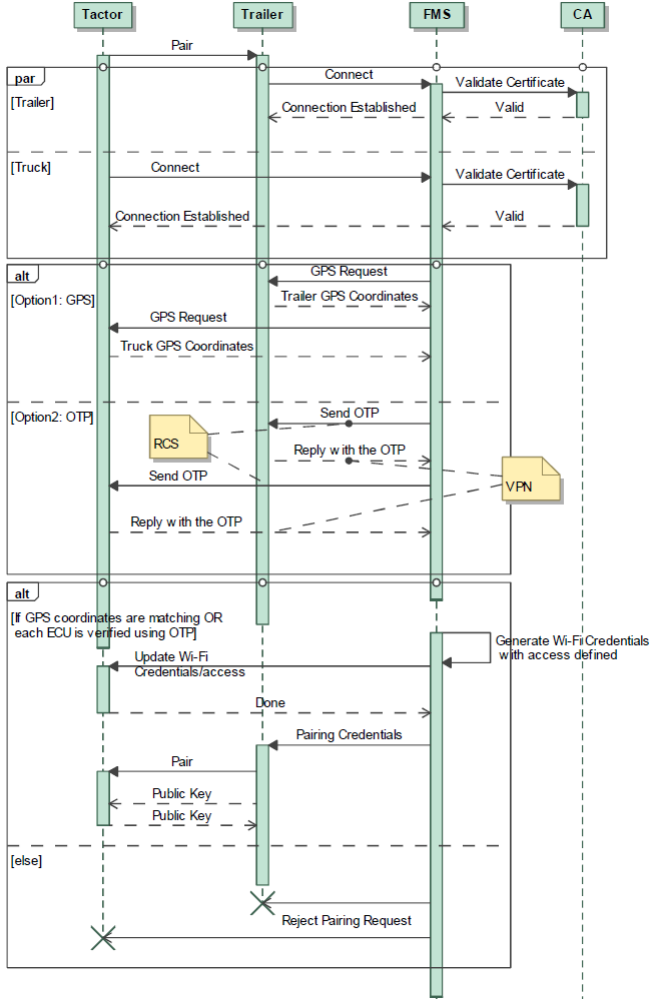

Figure 14. Authentication of the trailer and the tractor when connecting over the wireless harness using geo-location authentication or OTP.

**GPS Spoofing**

One of the concerns with using GPS coordinates as a geo-location authentication method during the tractor-trailer pairing process as show in figure 12 is GPS spoofing. The attackers use a transmitter that mimics the satellites to send fake GPS signals to the vehicle GPS receiver causing the vehicle GPS to report the attacker's desired coordinates. There are multiple approaches to detect and overcome GPS spoofing on the autonomous vehicle or autonomous tractor side such as dead-reckoning, prediction, sensor-fusion or Visual Positioning System. The trailer could leverage the GPS from a coupled trusted AT as a source of truth and use it to detect spoofing. Assuming a trusted AT $T$ with an accurate GPS and connected to a trailer $R$ and the distance between the AT GPS and the trailer GPS is $d$ with a maximum GPS error of $e_T$ and $e_R$ respectively. The series of coordinates $x$ and $y$ for the AT and the trailer are described as follows:

$$[(x_{T1_{t1}}, y_{T1_{t1}}), (x_{T2_{t2}}, y_{T2_{t2}}), \dots, (x_{Ti_{ti}}, y_{Ti_{ti}})] \quad (1)$$

$$[(x_{R1_{t1}}, y_{R1_{t1}}), (x_{R2_{t2}}, y_{R2_{t2}}), \dots, (x_{Ri_{ti}}, y_{Ri_{ti}})] \quad (2)$$

From (1) and (2) series, any coordinates with series position $n$ and timestamp $tn$ will be evaluated using the following criteria:

$$|x_{Tn_{tn}} - (x_{Rn_{tn}} + d)| \leq e_T + e_R \quad (3)$$

$$|y_{Tn_{tn}} - (y_{Rn_{tn}} + d)| \leq e_T + e_R \quad (4)$$

If (3) or (4) is False, a GPS spoofing or malfunction is occurring on the trailer side. An alternative method in case geo-location authentication is not possible is sending a OTP in a message over Rich Communication Services (RCS) from the cloud to each ECU as shown in figure 14. After the two ECUs are authenticated by the cloud over the VPN, for example, the cloud could send a code number to each ECU using a new different channel such as RCS which uses IP and encryption. When each ECU receive their OTP, they will send it again to the cloud over the VPN. Both of GPS and OTP could be used as part of a Multi-factor Authentication (MFA) process.

**Impact On the ECUs Lifecycle**

The AT and the trailer ECUs will be subject to new attack vectors due to the use of new interfaces and features such as Ethernet or wireless harness and cellular connection which will have an impact on the lifecycle and new design considerations. Based on the proposed changes to the trailer ABS and as suggested by [50], NHTSA [51], SAE J3101 [61], and SAE J3061 [62] new design consideration, development and lifecycle process are needed for the trailer ECU, especially when it comes to security such as following Security Development Lifecycle (SDL). Figure 15 shows a multi-layer cybersecurity concept to protect against the new attacks on the ECU.

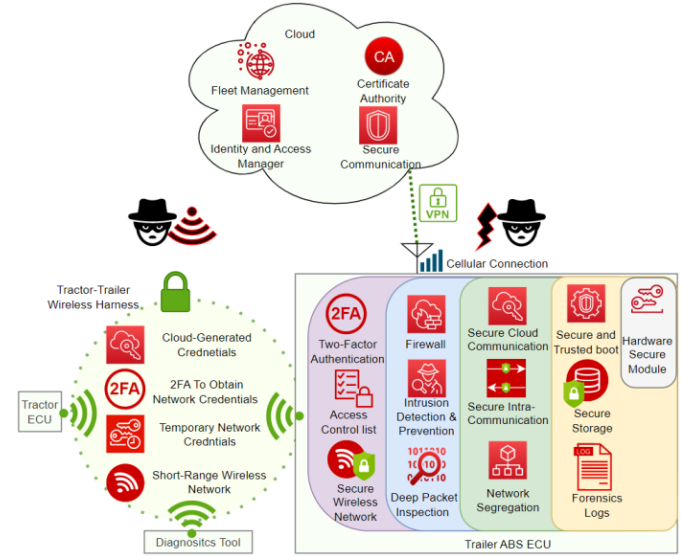

Figure 15. A Multi-layer security concept for the new trailer ABS ECU and the tractor ECU

Table 7 shows the impact on the lifecycle of the ECU and the new activities that needs to be taken into consideration.

Table 7. Main impact of the new features and interfaces on the trailer ABS lifecycle phases and technical processes per the V-model and INCOSE [63]

| Technical Process | Phase Impact |
|---|---|
| System Requirements Definition | → ABS Hardware and software security and performance requirements<br>→ Security requirements for the new communication channels (e.g., in-vehicle and cloud), hardware and software such as authorization, authentication, and data integrity and confidentiality.<br>→ Wireless connections requirements such as the wireless harness and the cellular connection performance and bandwidth<br>→ Design and cybersecurity risk assessment of adding Wi-Fi and cellular to the ABS.<br>→ Distributed development in the case of the tractor and the trailer from two different OEMs.<br>→ Requirements from different stakeholders (tractor and trailer OEMs) |
| System Design | → Design and security specification and the architecture of the new system including deployment, software, hardware, and cloud architecture.<br>→ System analysis including cost, technical risks, and effectiveness analysis<br>→ Cybersecurity analysis to configure the proper cybersecurity level for the system<br>→ Safety and cybersecurity by design |
| Implementation and Integration | → Secure hardware and software implementation<br>→ Integration, configuration, and testing of software and hardware components<br>→ Secure IT infrastructure<br>→ New procedures and training |
| Verification, Validation and Testing | → Scanning for vulnerabilities in the software and the hardware<br>→ Reverse Engineering, Fuzzing, and penetration testing<br>→ Conformance testing of the security functions and implementation<br>→ Testing of the wireless harness under different conditions including end-to-end data gatewaying<br>→ Features and cybersecurity integration testing<br>→ Software and hardware integration testing |
| Production, Operation, Maintenance and Updates | → Refined security assessment<br>→ Personnel Training and cybersecurity culture<br>→ Diagnostics over the wireless harness and the impact on the tools and protocols used.<br>→ Software updates process such as Over-the-air or over-the-wireless-harness updates<br>→ Fleet monitoring for security incidents and incident response<br>→ Pairing credentials handling and management |
| Disposal | → Disposal procedure and strategy<br>→ Secure disposal of the system and the data it contains |

## Identity And Access Management

FMS will be responsible for access control of the wireless harness network where it regulates the level of access for each Wi-Fi credentials. FMS will be able to add and update the access control list (ACL) in each ECU to define the users and groups for each entity connecting to the wireless harness networking using the provided credentials. For example, doing diagnostics over the wireless harness will be possible for both AT and the trailer using DoIP and the Wi-Fi

Page 10 of 13connection credentials for a technician with a diagnostic tool will have a different access level compared to the ECU pairing credentials. In both cases, credentials are managed and generated by FMS and stored and updated regularly within the ECU.

## Summary/Conclusions

For SAE level 4 and 5 autonomous tractors to drive in reverse, it needs additional autonomy sensors on the back of the trailer and current trailer ABS ECU cannot support autonomous tractor networking or have autonomy sensors connected due to the limitation in the computation, networking and architecture. We proposed a new trailer ABS ECU architecture that contain all of the existing features such as telematics, lights and GPS and uses automotive Ethernet or a wireless harness as the only communication link with the autonomous tractor in addition to using Named Data Networking. NDN is a new and promising networking architecture that could be standardized in the automotive industry to reduce complexity and have security by default in the data and interest packets. We discussed NDN and evaluated it against Data Distribution Service (DDS) and the experiment had position results. The test shows the NDN over TCP is an efficient protocol that is capable of meeting automotive communication requirements. We presented an automated tractor-trailer pairing method in addition to the security measures to authenticate each of them before pairing, in addition to the impact on different aspects of the lifecycle. Using Ethernet or a wireless harness and NDN for commercial trailer ABS ECU provides adequate resources for the operation of autonomous trucks and the expansion of its capabilities, and at the same time significantly reduces the complexities compared to when new features are added to legacy communication systems. Future work will include the networking and security specification for Named Data Networking when used between the trailer and the tractor and test and evaluation of the proposed system using twisted-pair automotive Ethernet and automotive ECUs and evaluation of the wireless harness concept.

## References

[1] SAE International, "Taxonomy and Definitions for Terms Related to Driving Automation Systems for On-Road Motor Vehicles," no. J3016_202104.

[2] P. Nyberg, "Stabilization, Sensor Fusion and Path Following for Autonomous Reversing of a Full-scale Truck and Trailer System," *Linköping University,* 2016.

[3] V. Josef, "Trailer parking assistant," in *Proceedings of the 16th International Conference on Mechatronics - Mechatronika 2014*, 2014.

[4] D. Parthasarathy, R. Whiton, J. Hagerskans and T. Gustafsson, "An in-vehicle wireless sensor network for heavy vehicles," *2016 IEEE 21st International Conference on Emerging Technologies and Factory Automation (ETFA),* pp. 1-8, 2016.

[5] J.-R. Lin, T. Talty and O. K. Tonguz, "An empirical performance study of Intra-vehicular Wireless Sensor Networks under WiFi and Bluetooth interference," *2013 IEEE Global Communications Conference (GLOBECOM),* pp. 581-586, 2013.

[6] M. Potdar and S. Wani, "Wireless Sensor Network in Vehicles," *SAE Technical Paper 2015-01-0241,* 2015.

[7] B. Shaer, D. L. Marcum, C. Becker, G. Gressett and M. Schmieder, "Wireless Blind Spot Detection and Embedded

## Contact Information

## Definitions/Abbreviations

| | |
|---|---|
| **AT** | Autonomous Tractor or Truck |
| **ECU** | Electronic Control Unit |
| **NDN** | Named Data Networking |
| **NFD** | Named Data Networking Forwarding Daemon |
| **DDS** | Data Distribution Service |
| **CAN** | Controller Area Network |
| **CAN FD** | Controller Area Network Flexible Data |
| **SoC** | System-on-Chip |

| | | | |
|---|---|---|---|
| **MCU** | Microcontroller | **AUTOSAR** | Automotive open system architecture |
| **TCP** | Transmission Control Protocol | **DCU** | Domain Controller Unit |
| **UDP** | User Datagram Protocol | **NMFTA** | National Motor Freight Traffic Association, Inc. |
| **RTPS** | Real-Time Publish Subscribe protocol | **RCS** | Rich Communication Service |
| **ACK** | Acknowledgement | **OTP** | One-Time Passcode |
| **ABS** | Anti-Lock Braking System | **ACL** | Access Control List |
| **GPS** | Global Positioning System | **FMS** | Fleet Management System |
| **CPU** | Central Processing Unit | **SAE** | Society of Automotive Engineers |
| **PLC** | Power Line Carrier | | |
| **AVB** | Audio Video Bridging | **VPN** | Virtual Private Network |
| **TSN** | Time-Sensitive Networking | **NHTSA** | National Highway Traffic Safety Administration |
| **SOME/IP** | Scalable service-Oriented MiddlewarE over IP | **SDL** | Security Development Lifecycle |
| **ADAS** | Advanced Driver-Assistance System | **CSR** | Certificate Signing Request |
| **UWB** | Ultra-wideband | | |
| **DoIP** | Diagnostics over IP | | |
| **UDS** | Unified Diagnostic Services | | |